\documentclass[twocolumn,showpacs,preprintnumbers,amsmath,amssymb,times,pre]{revtex4}

\usepackage{graphicx}
\usepackage{dcolumn}
\usepackage{bm}
\usepackage{amsmath}
\usepackage{color}

\begin{document}

\title{Promoting collective motion of self-propelled agents by distance-based influence}

\author{Han-Xin Yang$^{1}$}\email{hxyang01@gmail.com}

\author{Tao Zhou$^{2}$}\email{zhutou@ustc.edu}
\author{Liang Huang$^{3}$}\email{huangl@lzu.edu.cn}

\affiliation{$^{1}$Department of Physics, Fuzhou University, Fuzhou
350108, China\\$^{2}$Web Sciences Center, University of
Electronic Science and Technology of China, Chengdu 610054,
China\\$^{3}$Institute of Computational Physics and Complex Systems,
Lanzhou University, Lanzhou, Gansu 730000, China}

\begin{abstract}
We propose a dynamic model for a system consisting of self-propelled
agents in which the influence of an agent on another agent is
weighted by geographical distance. A parameter $\alpha$ is
introduced to adjust the influence: the smaller value of $\alpha$
means that the closer neighbors have stronger influence on the
moving direction. We find that there exists an optimal value of
$\alpha$, leading to the highest degree of direction consensus. The
value of optimal $\alpha$ increases as the system size increases,
while it decreases as the absolute velocity, the sensing radius and
the noise amplitude increase.
\end{abstract}

\date{\today}

\pacs{89.75.Fb, 45.50.-j, 05.45.Xt}

\maketitle

\section{Introduction}

The collective motion is a ubiquitous phenomenon in nature, examples
of which include traffic jams~\cite{traffic}, bird
flocks~\cite{bird0,bird1,bird2,bird3}, fish schools~\cite{fish},
insects swarms~\cite{insect1,insect2}, bacteria
colonies~\cite{bacteria1,bacteria2}, pedestrian
flows~\cite{flow0,flow1,flow2} and active granular
media~\cite{media}. In recent years, a variety of efforts have been
devoted to modeling the dynamic properties of
swarms~\cite{report,1,2,4,5,6,7,8,9,10,10.1,11,12}. In 1995, Vicsek
$et$ $al$. proposed a particularly simple but rich
model~\cite{vicsek}. In the Vicsek model (VM), some self-propelled
agents move with the same absolute velocity in a square-shaped cell
with the periodic boundary conditions. At each time step, every
agent updates its direction according to the average direction of
the agents' motion in its neighborhood. The neighborhood of an agent
$i$ is composed by agent $i$ itself and those agents who fall in a
circle of sensing radius that centered at the current position of
$i$. It has been demonstrated that all agents will converge to the
same direction on a macroscopic scale when the density of the system
is high and the noise is small enough~\cite{same}.

The VM and its variations have attracted much attention in the past
decade~\cite{onset,Huepe2,yang,li,bian,li2,tian,peng,gao1,Schubring}.
Gr\'{e}goire and Chat\'{e} found that the onset of collective motion
in the VM as well as in related models with and without cohesion is
always discontinuous~\cite{onset}. Huepe and Aldana studied
intermittency and clustering in the VM~\cite{Huepe2}. Yang $et$
$al$. considered a power-law distribution of sensing radius which
can enhance the convergence efficiency~\cite{yang}. Li $et$ $al$.
proposed an adaptive velocity model in which each agent not only
adjusts its moving direction but also adjusts its speed according to
the degree of direction consensus among its local
neighbors~\cite{li}. Tian $et$ $al$. discovered that there exists an
optimal view angle, leading to the fastest direction
consensus~\cite{tian}. Gao $et$ $al$. proposed a restricted angle
model that significantly improves the collective motion of
self-propelled agents~\cite{gao1}. Schubring and Ohmann proposed a
density-independent modification of the VM in which an agent
interacts with neighbors defined by Delaunay
triangulation~\cite{Schubring}. Peruani and B\"{a}r demonstrated
that the clustering statistics and the corresponding phase
transition to non-equilibrium clustering found in many experiments
and simulation studies with self-propelled particles with alignment
can be obtained by a simple kinetic model~\cite{Peruani}. Buscarino
$et$ $al$. found that the direction consensus can be improved by
long-range interactions~\cite{Buscarino1,Buscarino2}.

In the original VM, all agents in agent $i$'s neighborhood have the
same influence on the moving direction of agent $i$. However, due to
the existence of diversity in human society and animal world,
influences of different agents usually are not the same. It has been
shown that the heterogeneous influence plays an important role in
various dynamics, such as the formation of public
opinion~\cite{opinion1,opinion2} and the evolution of
cooperation~\cite{game}. In this paper, we propose a weighted VM,
where the influence of neighbor $j$ on agent $i$ is determined by
the geographical distance between the two agents. We set the weight
of neighbors to be an exponential function with a tunable parameter
$\alpha$. Interestingly, we find that there exists an optimal value
of $\alpha$, leading to the highest degree of direction consensus.
The effects of the moving speed, the sensing radius, the system size
and the noise amplitude on direction consensus of the system are
also studied.

The paper is organized as follows. In Sec.~\ref{sec:model}, we
introduce the distance-based influence model. Simulation and
discussion are given in Sec.~\ref{sec: results}. The paper is
concluded in Sec.~\ref{sec: conclusion}.

\section{The distance-based influence model}\label{sec:model}

We consider $N$ agents moving in the two-dimensional plane without
periodic boundary conditions~\cite{li,gao2}. Initially agents are
randomly distributed on a region of $L\times L$ rectangle with
random directions. Note that this rectangle does not represent the
boundary for motion, but only restricts the initial distribution of
positions of agents. Each agent has the same absolute velocity
$v_{0}$ and sensing radius $r$. At time $t$, the position of a
specific agent $i$ is updated according to
\begin{equation}
\textbf{x}_{i}(t+1)=\textbf{x}_{i}(t)+v_{0}e^{i\theta(t)}. \label{1}
\end{equation}
Its direction is updated as
\begin{equation}
e^{i\theta_{i}(t+1)}=e^{i\Delta\theta_{i}(t)}\frac{\sum_{j\in\Gamma_{i}(t)}W_{j}(t)e^{i\theta_{j}(t)}}{\parallel
\sum_{j\in\Gamma_{i}(t)}W_{j}(t)e^{i\theta_{j}(t)} \parallel},
\label{2}
\end{equation}
where $\Delta\theta_{i}\in[-\eta, \eta]$ denotes the white noise,
$e^{i\theta(t)}$ is a unit directional vector, $\Gamma_{i}(t)$ is
the set of neighbors of agent $i$ defined by the sensing radius $r$
at time step $t$, and the weight $W_{j}(t)$ denotes the influence of
the neighbor $j$ of agent $i$ at time step $t$. We define $W_{j}(t)$
as
\begin{equation}
W_{j}(t)=e^{\alpha\cdot d_{ij}(t)}, \label{3}
\end{equation}
where $d_{ij}(t)=\|\textbf{x}_{j}(t)-\textbf{x}_{i}(t)\|$ is the
distance between agent $i$ and $j$ at time step $t$, and $\alpha$ is
a tunable parameter. For $\alpha>0 (<0)$, the farther (closer)
neighbors have larger weight of influence. When $\alpha=0$, our model
is the same as the standard VM, where each agent in the system has
the same weight.

\section{Simulation and discussion}\label{sec: results}

In all the following simulations, we set $L=10$.

\begin{figure}
\begin{center}
\scalebox{0.42}[0.42]{\includegraphics{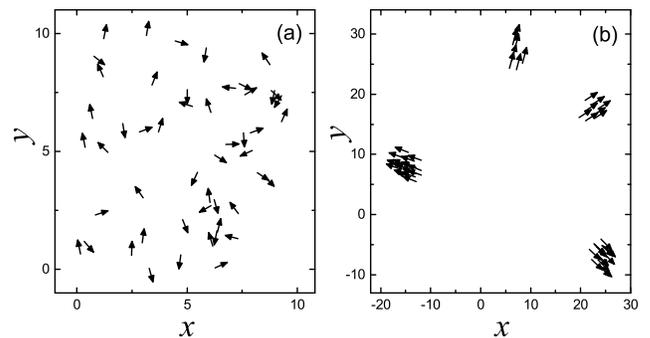}}
\caption{Snapshots of locations and velocities (a) in the initial
configuration, and (b) at the time step $t=100$. The arrows show the
direction of motion of the agents. The parameters are set as $L=10$,
$N=50$, $r=2$, $v_{0}=0.2$, and $\alpha=0$.} \label{fig1}
\end{center}
\end{figure}

We first consider the case in which the noise is zero ($\eta=0$).
From the perspective of complex network theory, the topology of
self-propelled agent system can be expressed as a temporal
network~\cite{petter}. At time $t$, each agent is represented by a
node and an edge between agent $i$ and $j$ is established if the
distance between them is shorter than the sensing radius $r$. A
cluster is a subgraph in which any two nodes are connected to each
other by paths running along edges of the network. In the case of
zero noise, after a period of evolution, agents aggregate into
different moving polar clusters. All agents within a moving polar
cluster move in the same direction, as shown in Fig.~\ref{fig1}.

\begin{figure}
\begin{center}
\scalebox{0.4}[0.4]{\includegraphics{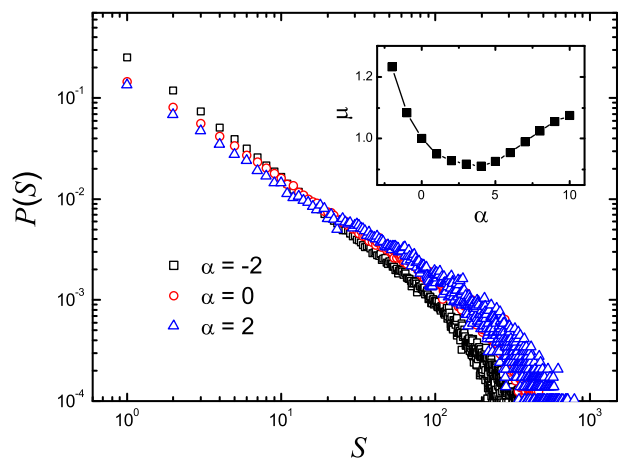}} \caption{(Color
online) The distribution of cluster sizes $P(S)$ for different
values of $\alpha$. It is found that $P(S)\sim S^{-\mu}$. The inset
shows that the exponent $\mu$ as a function of $\alpha$. The
absolute velocity $v_{0}=1$, the sensing radius $r=1.6$ and the
system size $N=1000$. Each data point results from an average over
$10^{4}$ different realizations.} \label{fig2}
\end{center}
\end{figure}

Following the previous studies~\cite{bacteria2,Peruani0}, we examine
the cluster size distribution. The results are shown in
Fig.~\ref{fig2}. We see that the probability $P(S)$ that a cluster
with size $S$ decays as a power law, following $P(S)\sim S^{-\mu}$.
The inset of Fig.~\ref{fig2} shows the dependence of $\mu$ on
$\alpha$ when the absolute velocity $v_{0}=1$ and the sensing radius
$r=1.6$. We find that the exponent $\mu$ is minimum at
$\alpha\approx4$. Since smaller $\mu$ indicates a higher probability
of large clusters, $\alpha\approx4$ is an optimal value for the
system to have large clusters. The value of $\mu$ is found to be in
the range $0.8 <\mu< 1.5$ in line with the previous experiments and
simulations~\cite{bacteria2,Peruani,Peruani1}.

Following the previous studies~\cite{yang,li,gao1}, we measure the
degree of direction consensus by the relative size $s_{1}$ of the
largest cluster, which is defined as the ratio of the number of
agents within the largest cluster to the total number of agents.
Note that $0\leq s_{1}\leq1$. A larger value of $s_{1}$ indicates a
higher degree of direction consensus.

\begin{figure}
\begin{center}
\scalebox{0.4}[0.4]{\includegraphics{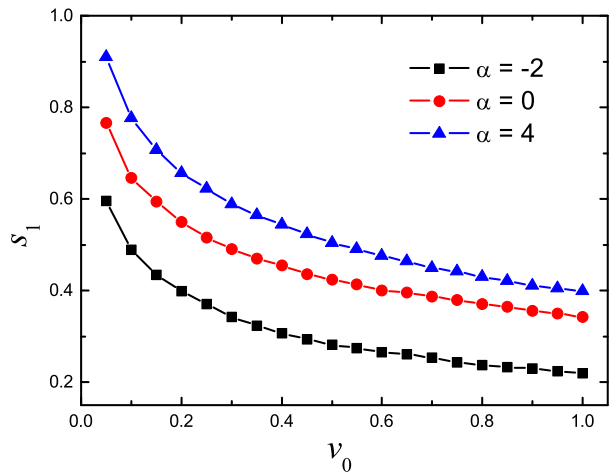}} \caption{(Color
online) The relative size $s_{1}$ of the largest cluster in the
steady state, as a function of the absolute velocity $v_{0}$ for
different values of $\alpha$. The sensing radius $r=1.6$ and the
system size $N=1000$. Each data point results from an average over
1000 different realizations.} \label{fig3}
\end{center}
\end{figure}

\begin{figure}
\begin{center}
\scalebox{0.4}[0.4]{\includegraphics{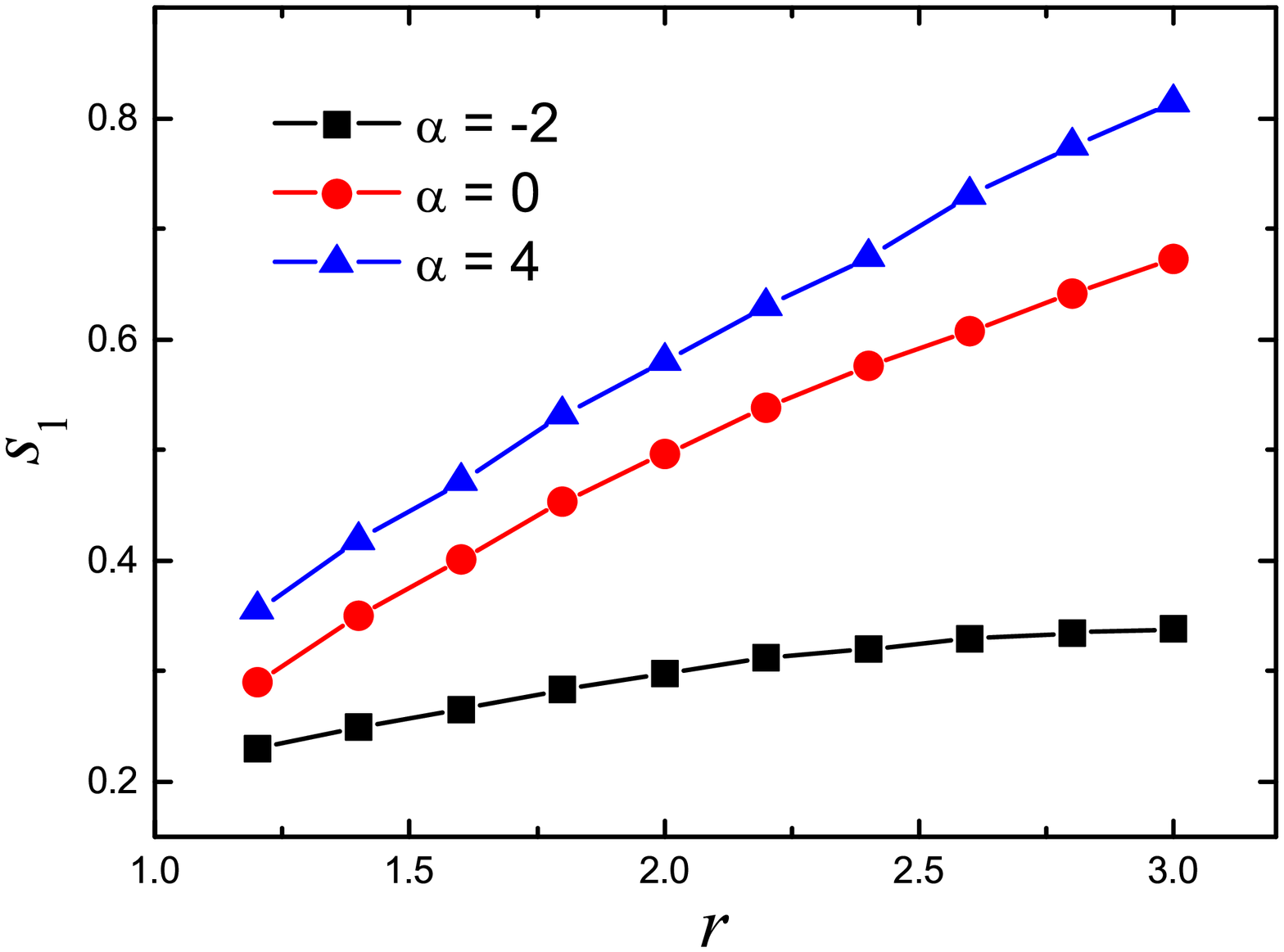}} \caption{(Color
online) The relative size $s_{1}$ of the largest cluster in the
steady state, as a function of the sensing radius $r$ for different
values of $\alpha$. The absolute velocity $v_{0}=0.6$ and the system
size $N=1000$. Each data point results from an average over 1000
different realizations. }\label{fig4}
\end{center}
\end{figure}

\begin{figure}
\begin{center}
\scalebox{0.4}[0.4]{\includegraphics{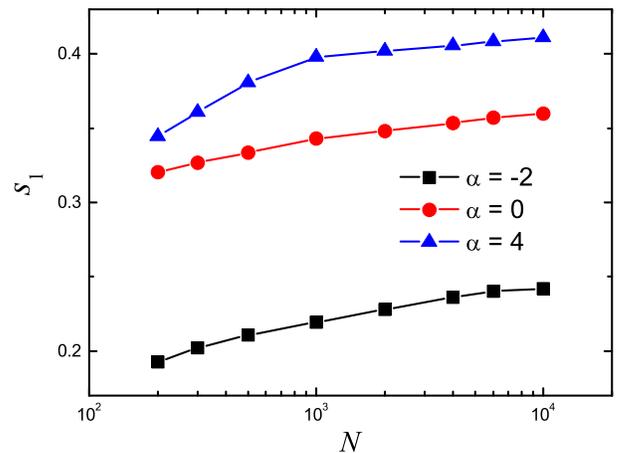}} \caption{(Color
online) The relative size $s_{1}$ of the largest cluster in the
steady state, as a function of the system size $N$ for different
values of $\alpha$. The absolute velocity $v_{0}=1$ and the sensing
radius $r=1.6$. Each data point results from an average over 1000
different realizations. }\label{new5}
\end{center}
\end{figure}

Figure~\ref{fig3} shows that $s_{1}$ as a function of the absolute
velocity $v_{0}$ for different values of $\alpha$. From
Fig.~\ref{fig3}, we can see that for any given value of $\alpha$,
$s_{1}$ decreases monotonically as $v_{0}$ increases.
Figure~\ref{fig4} plots $s_{1}$ as a function of the sensing radius
$r$ for different values of $\alpha$. In Fig.~\ref{fig4}, one can
observe that for any given value of $\alpha$, $s_{1}$ increases with
$r$. Figure~\ref{new5} depicts $s_{1}$ as a function of the system
size $N$ for different values of $\alpha$. From Fig.~\ref{new5}, we
find that for any given value of $\alpha$, $s_{1}$ increases with
$N$.

\begin{figure}
\begin{center}
\scalebox{0.4}[0.4]{\includegraphics{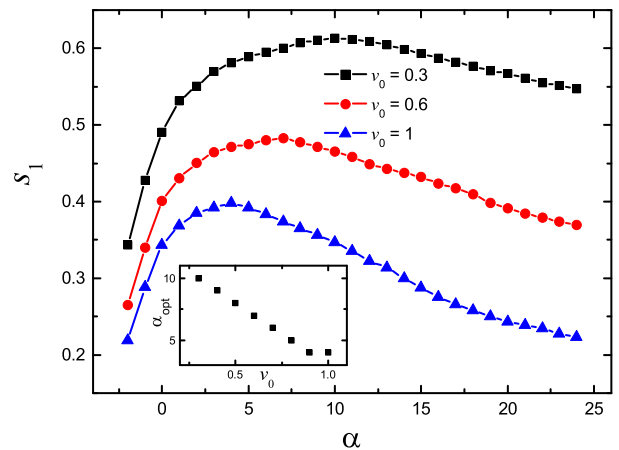}} \caption{(Color
online) The relative size $s_{1}$ of the largest cluster in the
steady state, as a function of $\alpha$ for different values of the
absolute velocity $v_{0}$. The inset shows the optimal value of
$\alpha$, $\alpha_{opt}$, as a function of $v_{0}$. The sensing
radius $r=1.6$ and the system size $N=1000$. Each data point results
from an average over 1000 different realizations.} \label{fig5}
\end{center}
\end{figure}

Figure~\ref{fig5} shows the dependence of $s_{1}$ on $\alpha$ for
different values of the absolute velocity $v_{0}$. In
Fig.~\ref{fig5}, one can observe a resonance-like behavior. For a
given value of $v_{0}$, there exists an optimal value of $\alpha$,
hereafter denoted by $\alpha_{opt}$, leading to the maximum $s_{1}$.
It is interesting to note that, for the absolute velocity $v_{0}=1$
and the sensing radius $r=1.6$, $\alpha_{opt}\approx4$ not only
corresponds to the maximum $s_{1}$, but also results in the minimum
$\mu$ (see the inset of Fig.~\ref{fig2}). Intuitively, at this
point, a minimum $\mu$ corresponds to a distribution $P(S)$ that
decreases most slowly as $S$ increases, indicating a larger
probability that a large cluster emerges. This infers that the
minimum of $\mu$ and the maximum of $s_{1}$ occur at the same point
of $\alpha_{opt}\approx4$. From the inset of Fig.~\ref{fig5}, we
find that $\alpha_{opt}$ decreases from 10 to 4 as $v_{0}$ increases
from 0.3 to 1. Since larger $\alpha$ indicates more influence from
distant nodes, this means that as $v_{0}$ increases, direction
consensus is better preserved by strengthening the influence from
closer agents.

\begin{figure}
\begin{center}
\scalebox{0.4}[0.4]{\includegraphics{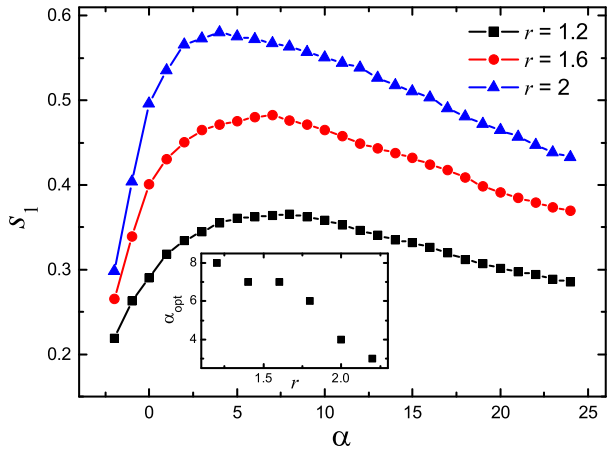}} \caption{(Color
online) The relative size $s_{1}$ of the largest cluster in the
steady state, as a function of $\alpha$ for different values of the
sensing radius $r$. The inset shows the optimal value of $\alpha$,
$\alpha_{opt}$, as a function of $r$. The absolute velocity
$v_{0}=0.6$ and the system size $N=1000$. Each data point results
from an average over 1000 different realizations. }\label{fig6}
\end{center}
\end{figure}

\begin{figure}
\begin{center}
\scalebox{0.4}[0.4]{\includegraphics{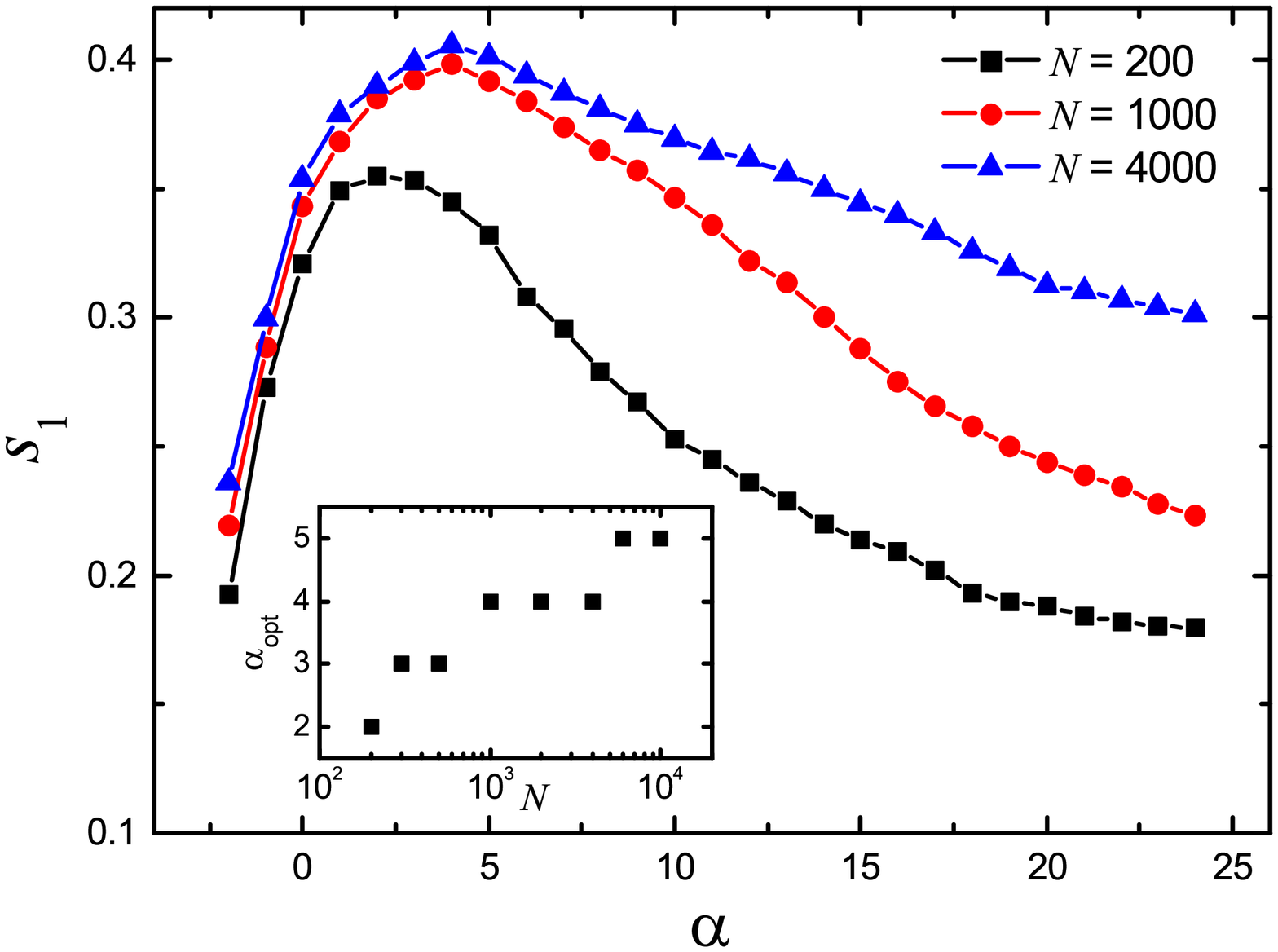}} \caption{(Color
online) The relative size $s_{1}$ of the largest cluster in the
steady state, as a function of $\alpha$ for different values of the
system size $N$. The inset shows the optimal value of $\alpha$,
$\alpha_{opt}$, as a function of $N$. The absolute velocity
$v_{0}=1$ and the sensing radius $r=1.6$. Each data point results
from an average over 1000 different realizations. }\label{new8}
\end{center}
\end{figure}

Figure~\ref{fig6} shows the dependence of $s_{1}$ on $\alpha$ for
different values of the sensing radius $r$. From Fig.~\ref{fig6}, we
find that for a given value of $r$, there exists an optimal value of
$\alpha$ leading to the maximum $s_{1}$. The inset of
Fig.~\ref{fig6} displays that $\alpha_{opt}$ decreases from 8 to 3
as $r$ increases from 1.2 to 2.4. Figure~\ref{new8} shows the
dependence of $s_{1}$ on $\alpha$ for different values of the system
size $N$. From Fig.~\ref{new8}, one can see that for a given value
of $N$, there exists an optimal value of $\alpha$ resulting in the
maximum $s_{1}$. The inset of Fig.~\ref{new8} displays that
$\alpha_{opt}$ increases from 2 to 5 as $N$ increases from 200 to
10000.

\begin{figure}
\begin{center}
\scalebox{0.4}[0.4]{\includegraphics{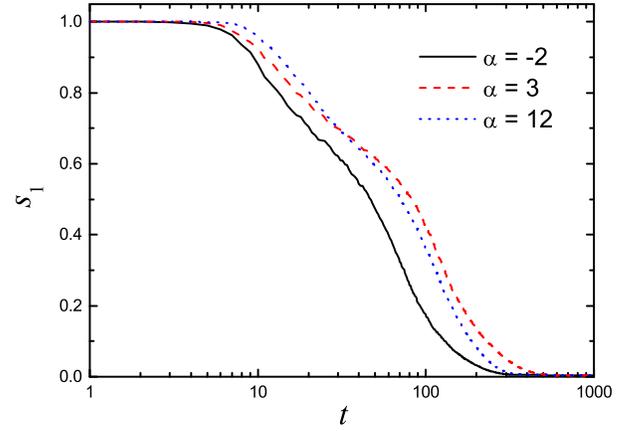}} \caption{(Color
online) The relative size $s_{1}$ of the largest cluster as a
function of the time step $t$ for different values of $\alpha$. The
absolute velocity $v_{0}=0.6$, the sensing radius $r=1.6$, the
system size $N=1000$ and the noise amplitude $\eta=0.6$. Each curve
results from an average over 1000 different
realizations.}\label{fig7}
\end{center}
\end{figure}

The resonance-like behavior of the dependence of $s_1$ on $\alpha$
can be understood as follows. For very small values of $\alpha$, an
agent is mostly influenced by its closest neighbors and the active
interaction radius is much shorter than the sensing radius $r$. As
indicated by the $\alpha=-2$ case in Fig.~\ref{fig4}, $s_1$
increases very slowly with $r$. Previous studies have shown that the
direction consensus can be improved by adding long-range
interactions~\cite{Buscarino1,Buscarino2,zhang}. Thus it is
beneficial if distant neighbors are given a strong influence.
However, when $\alpha$ is too large that an agent is mostly
influenced by the farthest neighbor, the interaction between other
neighbors would be so weak that is not able to maintain the coherent
motion between them, which in turn also disrupts direction
consensus. A similar effect has been observed in percolation
issues~\cite{huang}. Therefore, there must be an intermediate value
of $\alpha$ that optimizes direction consensus.

Next, we study the case in which the noise amplitude $\eta$ is
nonzero. Figure~\ref{fig7} shows that $s_{1}$ as a function of the
time step $t$ for different values of $\alpha$ when $\eta=0.6$. From
Fig.~\ref{fig7}, we see that $s_{1}$ decreases to 0 as time evolves,
indicating that all the agents will disperse away without any
apparent cluster in the noisy environment. From Fig.~\ref{fig7}, one
can also observe that the value of $\alpha$ affects the process of
dispersion.

\begin{figure}
\begin{center}
\scalebox{0.4}[0.4]{\includegraphics{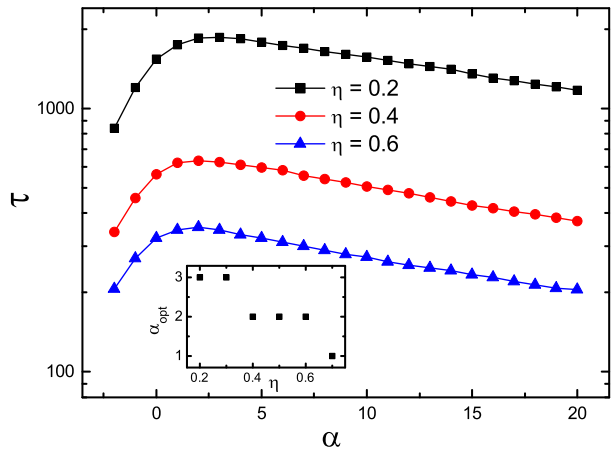}} \caption{(Color
online) The transient time $\tau$ as a function of $\alpha$ for
different values of the noise amplitude $\eta$. The inset displays
the optimal value of $\alpha$, $\alpha_{opt}$, as a function of
$\eta$. The absolute velocity $v_{0}=0.6$, the sensing radius
$r=1.6$ and the system size $N=1000$. Each data point results from
an average over 1000 different realizations. }\label{fig8}
\end{center}
\end{figure}

To quantify the speed of dispersion in the noisy case, we study the
transient time $\tau$, which is defined as the time when $s_{1}$
first below a certain value $\epsilon$. We have checked that
qualitative results are invariant when $\epsilon$ is small enough.
In this paper we take $\epsilon=0.02$. Figure~\ref{fig8} shows the
transient time $\tau$ as a function of $\alpha$ for different values
of the noise amplitude $\eta$. One can see that $\tau$ can be
maximized by an optimal value of $\alpha$. This means the process of
dispersion can be best slowed down by fine tuning the value of
$\alpha$. Moreover, the inset of Fig.~\ref{fig8} displays that the
optimal value of $\alpha$, $\alpha_{opt}$ decreases from 3 to 1 as
the noise amplitude $\eta$ increases from 0.2 to 0.7.

\begin{figure}
\begin{center}
\scalebox{0.4}[0.4]{\includegraphics{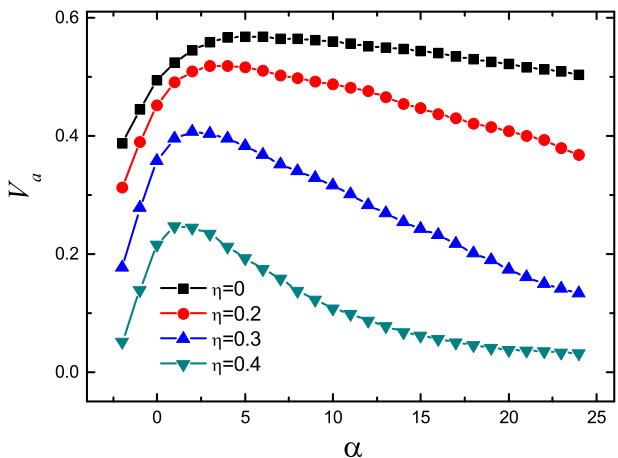}} \caption{(Color
online) The order parameter $V_{a}$ as a function of $\alpha$ for
different values of the noise amplitude $\eta$. The absolute
velocity $v_{0}=0.6$, the sensing radius $r=1.6$ and the system size
$N=1000$. Here $V_{a}$ is calculated at the time step $t=500$. Each
data point results from an average over 1000 different realizations.
}\label{fig9}
\end{center}
\end{figure}

In the above studies, the degree of direction consensus of a system
is quantified by the relative size of the largest cluster. We have
checked that the qualitative results hold unchanged if we use an
order parameter to measure the degree of direction consensus, as
defined by~\cite{vicsek}
\begin{equation}
V_{a}=\frac{1}{N}\parallel\sum_{i=1}^{N}e^{i\theta_{i}(t)}\parallel,
\quad 0\leq V_{a}\leq1.
\end{equation}
A larger value of $V_{a}$ indicates a better consensus.
Figure~\ref{fig9} shows the dependence of $V_{a}$ on $\alpha$ for
different values of the noise amplitude $\eta$. From
Fig.~\ref{fig9}, one can see that for a given value of $\eta$,
$V_{a}$ is maximized by an optimal value of $\alpha$. Besides, we
find that $V_{a}$ decreases as $\eta$ increases when $\alpha$ is
fixed.

\section{Conclusion}\label{sec: conclusion}

In conclusion, we have introduced a weight based on the geographical
distance into the original Vicsek model. A tunable parameter
$\alpha$ is used to govern the weight. The direction of each agent
is updated by the weighted average directions of its neighbors. In
the noiseless case, self-propelled agents will aggregate into
different clusters and the size distribution of clusters exhibits a
power-law form. We find that direction consensus is enhanced when
the absolute velocity is small, the sensing radius or the system
size is large. Interestingly, there exists an optimal value of
$\alpha$ that yields the highest level of direction consensus. In
the noisy case, agents gradually disperse away and each agent moves
lonely. This process of dispersion can be best slowed down by
optimizing $\alpha$. The value of optimal $\alpha$ depends on the
absolute velocity, the sensing radius, the system size and the noise
amplitude. For all kinds of cases, the optimal value of $\alpha$ is
always positive, indicating that a suitably strong influence of
distant neighbors can best enhance the direction conseusus of the
system.

Note that recently Gao $et$ $al$. proposed another weighted Vicsek
model~\cite{gao2}, where the weight of each agent is determined by
its number of neighbors. They found that the system becomes easier
to achieve direction consensus when agents with more neighbors are
allocated greater weight. In the weighted model proposed by Gao $et$
$al$., agents are assumed to be able to obtain global information,
i.e., an agent not only knows the number of its neighbors, but also
knows the number of neighbors' neighbors. In our model, an agent
only gets local information, i.e., the distance between it and its
neighbor. All together Ref.~\cite{gao2} and our work provide a
deeper understanding of the impact of heterogeneous influence on
collective motion.

\begin{acknowledgments}
This work was supported by the National Natural Science Foundation
of China (Grants No. 11222543, 11135001, 11375074, and 91024026),
the Program for New Century Excellent Talents in University under
Grant No. NCET-11-0070, the Natural Science Foundation of Fujian
Province of China (Grant No. 2013J05007), and the Research
Foundation of Fuzhou University (Grant No. 0110-600607).
\end{acknowledgments}

\end{document}